\documentclass{aa}

\usepackage[mathscr]{euscript}
\usepackage{times}
\usepackage{txfonts}
\usepackage{xcolor}

\providecommand{\dif}{\mathrm{d}}
\providecommand{\sound}{c_{\mathrm{s},0}}
\providecommand{\en}{\mathscr{E}}
\providecommand{\pot}{\mathscr{U}}

\providecommand{\rg}{r_{\mathrm{G}}}

\graphicspath{{mat/}}

\usepackage{hyperref}
\newcommand{\orcid}[1]{\href{https://orcid.org/#1}{\includegraphics[width=8pt]{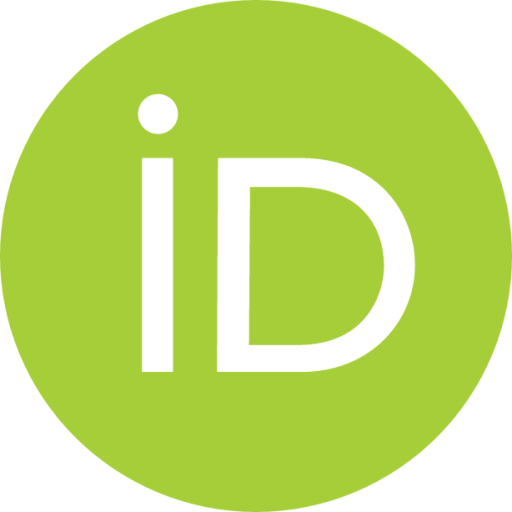}}}

\usepackage{comment}

\begin{document}

\title{Accretion tori around rotating neutron stars}
\subtitle{Paper I: Structure, shape and size} 
\titlerunning{Structure, shape and size of tori around NSs}

\author{
    \orcid{0000-0002-4193-653X}Monika Matuszkov\'a\inst{\ref{inst1},\ref{inst2}}
    \and
    \orcid{0000-0003-3958-9441}Gabriel T\"{o}r\"{o}k\inst{\ref{inst1}
    }
    \and
    \orcid{0000-0003-0826-9787}Debora Lan\v{c}ov\'a\inst{\ref{inst1}}
    \and
    \orcid{0000-0002-0930-0961}Kate\v{r}ina Klimovi\v{c}ov\'a\inst{\ref{inst1}}
    \and
    \orcid{0000-0002-7635-4839}Ji\v{r}\'{\i} Hor\'{a}k\inst{\ref{inst2}}
    \and\\
    \orcid{0000-0001-9635-5495}Martin Urbanec\inst{\ref{inst1}}
    \and
    \orcid{0009-0000-7736-6180}Eva \v{S}r\'{a}mkov\'{a}\inst{\ref{inst1}}
    \and
   \orcid{0000-0001-5755-0677}Odele Straub\inst{\ref{inst3},\ref{inst4}}
    \and
    \orcid{0000-0002-4480-5914}Gabriela Urbancov\'a\inst{\ref{inst1}}
    \and
    \orcid{0000-0002-5760-0459}Vladim\'{\i}r Karas\inst{\ref{inst2}}
}
\institute{
    Research Centre for Computational Physics and Data Processing, Institute of Physics, Silesian University in Opava, 
    Bezru\v{c}ovo n\'am.~13, CZ-746\,01 Opava, Czech Republic
    \label{inst1}
    \and
    Astronomical Institute of the Czech Academy of Sciences, Bo\v{c}n\'{\i} II 1401, CZ-14100 Prague, Czech Republic \label{inst2}
    \and
    ORIGINS Excellence Cluster, Boltzmannstr. 2, 85748 Garching, Germany
    \label{inst3}
    \and Max Planck Institute for Extraterrestrial Physics, Gießenbachstra{\ss}e 1, 85748 Garching, Germany 
    \label{inst4}
}
          
\authorrunning{Matuszkov\'a et al.}

\date{Received / Accepted}

\abstract{
We present a full general relativistic analytic solution for a radiation-pressure supported equilibrium fluid torus orbiting a rotating neutron star (NS). Previously developed analytical methods are thoroughly applied in the Hartle-Thorne geometry, including the effects of both the NS's angular momentum and quadrupole moment. The structure, size and shape of the torus are explored, focusing especially on the critically thick solution -- the cusp tori. For the astrophysically relevant range of NS parameters, we examine how our findings differ from those obtained for the Schwarzschild spacetime. The solutions for rotating stars display signatures of the interplay between relativistic and Newtonian effects where the impact of NS angular momentum and quadrupole moment are almost counterbalanced at the given radius. Nevertheless, the spacetime parameters still strongly influence the size of tori, which can be shown in a coordinate-independent way.  Finally, we discuss the importance of the size of the central NS, determining whether or not the surrounding torus may exist. We provide a set of tools in a Wolfram Mathematica code, which poses a basis allowing for a further investigation of the impact of the  NSs’ superdense matter equation of state on the spectral and temporal behaviour of accretion tori.
} 

\keywords  {
    Stars: neutron~--~Accretion, accretion disks}

\maketitle

\section{Introduction}
\label{sec:intro}

In the close vicinity of black holes (BHs) and neutron stars (NSs), accretion processes lead to the formation of discs emitting a strong radiative power that allows us to identify the nature of these compact objects and study their properties \citep[see e.g.][]{2002apa..book.....F,2008bhad.book.....K}. In the inner accretion region, the strong gravitational effects dominate, which affect the observed variability and spectra. These are directly dependent only on the geometry of spacetime in the case of BHs, while in the case of NSs, the nature of the superdense matter of which they consist plays a role as well \citep[see, e.g.][]{lewin-etal:2010:book,2007PhR...442..109L}.

In the last decades, several complementary approaches to investigate the properties of accreting BHs and NSs based on observed X-ray spectra and timing behaviour have been developed and gained popularity. Promising methods include, for example, fitting the X-ray spectral continuum or relativistically broadened iron lines \citep[][]{rey-fab:2000,2006csxs.book..157M,2008ApJ...676..549S,2011blho.book..252M,2003MNRAS.340L..28F,rey-fab:2008, 2009ApJ...701L..83S,2010ApJ...718L.117S,cac-etal:2008,pir-etal:2012,mil-etal:2020}. Within the time domain, the interpretation of observations of quasiperiodic oscillations (QPOs) and related proposed models are often debated \citep[e.g., ][]{wag:1999, kat:2001, ste-vie:1998a,cad-etal:2008,2001A&A...374L..19A,rez-etal:2003,2003MNRAS.344L..37R,bur-etal:2004:APJ,2006CQGra..23.1689A,2006MNRAS.369.1235B, Ingram+Done:2010,2018MNRAS.474.3967D,mis-etal:2017:}. 
Stationary disc models are commonly assumed in many approaches investigating BH and NS properties, although it is not quite clear whether, in real astrophysical situations, the dynamical timescales of the studied phenomena are shorter than the viscous and thermal timescales \citep[][]{2009A&A...498..471Q,2011blho.book..252M}.

A particular class of simplified stationary solutions of accretion flows in geometrically thick discs are the radiation-pressure-supported equilibrium fluid tori, often denoted as the Polish doughnuts \citep[e.g.,][]{Kozlowski-etal:1978:,1978A&A....63..221A,pac-abr:1982}, which have been intensely explored and utilised within various works assuming rotating BHs \cite[][]{2001A&A...374L..19A,2005A&A...436....1T,2017A&A...607A..68G,2020ApJ...895...61R,2020MNRAS.491..417R}. Here, we consider tori around rotating NSs and, for the first time, apply previously developed analytical methods to the case of the Hartle-Thorne geometry, including the effects of both the NS’s angular momentum and quadrupole moment.

We explore the structure, size and shape of the tori around rotating NSs and examine differences from the case of Schwarzschild spacetime while focusing especially on the critically thick solution – the cusp tori. Our study provides a basis for further astrophysical applications. These can be represented, for instance, by studying tori oscillations in relation to QPOs observed in NS systems, as discussed in our follow-up paper \citep[Paper II,~][]{paperII-key}.

The present paper (Paper I) is organised as follows. Sect.~\ref{ssec:HT} introduces the adopted spacetime description. In Sect.~\ref{Section:tori}, we build on earlier research on accretion tori and carry out a comprehensive analytical solution describing accretion tori around rotating NSs. This solution is then further explored within the astrophysically relevant range of spacetime parameters in Sect.~\ref{section:astro}. Finally, in Sect.~\ref{Section:conclusions}, we demonstrate the use of our formulae on various astrophysically relevant examples. We provide the reader with a short summary and state our main concluding remarks.

\section{Neutron star spacetime geometry}
\label{ssec:HT}

We describe the spacetime surrounding axisymmetric NS by the line element expressed in the Schwarzschild coordinates, $\{t, r, \theta, \varphi\}$,
\begin{align}
    \label{eq:stel}
	\dif s^2 = g_{tt} \dif t^2 + 2 g_{t\varphi} \dif t \dif \varphi + g_{rr} \dif r^2 + g_{\theta\theta} \dif \theta^2 + g_{\varphi\varphi} \dif \varphi^2.
\end{align}
We use units where $c = G = 1$ with $c$ being the speed of light and $G$ the gravitational constant. To measure distance, we use $\rg = GM/c^2$. The $(-\,+\,+\,+)$ metric signature is employed throughout the paper.

\subsection{The Hartle-Thorne geometry}

\label{section:metric}

The exterior solution of the Hartle-Thorne metric is characterised by three parameters: the gravitational mass $M$, angular momentum $J$, and quadrupole moment $Q$ of the star, while dimensionless forms of the angular momentum and the quadrupole moment, $j=J/M^2$ and $q=Q/M^3$, are usually used. 

The metric functions take the form \citep{2003gr.qc....12070A}\footnote{Note the misprints in the original paper.}:
\begin{eqnarray} \nonumber
	g_{tt} = &&-\left( 1- \frac{2M}{r} \right)\phantom{^{-1}} \left[ 1 + j^2 F_1 \left(r\right) + q F_2 \left(r\right) \right], \\ 
	\nonumber g_{rr} = && \phantom{-}\left( 1- \frac{2M}{r} \right)^{-1} \left[ 1 + j^2 G_1 \left(r\right) - q F_2 \left(r\right) \right] , \\ \nonumber
	g_{\theta\theta} = && \phantom{-}\left[ 1 + j^2 H_1 \left(r\right) + q H_2 \left(r\right) \right]r^2 , \\ \nonumber
	g_{\varphi\varphi} = && \phantom{-}\left[ 1 + j^2 H_1 \left(r\right) + q H_2 \left(r\right) \right] r^2\sin^2 \theta, \\ 
	g_{t\varphi} = && - \frac{2M^2}{r} j \sin^2 \theta , 
\end{eqnarray}
where, making use of the $u = \cos \theta$ substitution, one may write 
\begin{eqnarray} \nonumber
	F_1\left(r\right) = && - [ 8Mr^4(r-2M) ]^{-1}\\ 
	&& \nonumber  [ u^2 ( 48M^6 - 8M^5r - 24M^4r^2 - 30M^3r^3 - 60M^2r^4    \\  \nonumber  && +  135Mr^5   - 45r^6 ) 
	 +  (r-M) ( 16M^5 + 8M^4r \\ \nonumber  &&- 10M^2r^3 - 30Mr^4 + 15r^5 ) ] + A_1(r), \\ \nonumber 
	F_2\left(r\right) = && \left[ 8Mr (r-2M) \right]^{-1} \nonumber  \\ \nonumber &&
	[ 5 (3u^2 - 1) (r-M) ( 2M^2 + 6Mr - 3r^2 ) ] - A_1\left(r\right), \nonumber  \\ 
	G_1\left(r\right) = && [ 8Mr (r-2M) ]^{-1} \nonumber  \\ \nonumber &&
	[ ( L\left(r\right) - 72M^5r ) - 3u^2 ( L\left(r\right) - 56M^5r ) ] - A_1\left(r\right),  \\ 
	L\left(r\right) = && 80 M^6 + 8M^4r^2 + 10M^3r^3 + 20M^2r^4 - 45Mr^5 \nonumber \\ \nonumber && + 15r^6, \\
	A_1\left(r\right) = && \frac{15 (r^2-2M) (1-3u^2)}{16M^2} \ln \left( \frac{r}{r-2M} \right), \nonumber  \\ \nonumber
	H_1\left(r\right) = && ( 8Mr^4 )^{-1} ( 1-3u^2 ) ( 16M^5 + 8M^4r - 10M^2r^3 \\ \nonumber && + 15Mr^4 + 15r^5 )  + A_2\left(r\right), \nonumber  \\
	H_2\left(r\right) = && (8Mr)^{-1} 5 (1-3u^2) ( 2M^2 - 3Mr - 3r^2 ) - A_2\left(r\right), \nonumber  \\ 
	A_2\left(r\right) = && \frac{15 (r^2-2M) (3u^2-1)}{16M^2} \ln \left( \frac{r}{r-2M} \right). 
\end{eqnarray}

The Hartle-Thorne metric coincides with the Schwarzschild metric for $j=0$ and $q=0$. Substituting $j=a/M$ and $q=(a/M)^2$, and performing a coordinate transformation into the Boyer-Lindquist coordinates \citep{2003gr.qc....12070A},
\begin{align}
	r_{\mathrm{BL}} &= r - \frac{a^2}{2r^3} \left[ \left(r+2M\right) \left(r-2M\right) + u^2 \left(r-2M\right) \left(r+3M\right) \right], 
	\\ 
	\theta_{\mathrm{BL}} &= \theta - \frac{a^2}{2r^3} \left(r+2M\right) \cos \theta \sin \theta,
\end{align}
we obtain Kerr geometry expanded up to the second order in the dimensionless angular momentum.

\subsection{Spacetime parameters, NS equation of state, and quadrupole parameter}
\label{section:range}

A thorough discussion of the relevance of the Hartle-Thorne geometry for the calculations of orbital motion and QPO models is presented in \cite{2019ApJ...877...66U}. For our purposes, we provide here a brief overview of the acceptable parameter ranges that are implied by the up-to-date NS equation of state (EoS). Conservative estimates of the NS mass in accreting sources range from $1.4M_\odot$ to $2.5\,M_\odot$\footnote{Here, $M_\odot$ denotes the solar mass.} \citep{1996ApJ...469L...9S,Lattimer+Prakash:2001}. The NS's dimensionless angular momentum can reach a maximum of approximately $j_{\mathrm{max}}~\sim~0.7$ \citep{2011ApJ...728...12L}.

In relation to the NS oblateness, it is useful to consider the quadrupole parameter $q/j^2$. Although the stellar oblateness can, in principle, be non-zero for a non-rotating star, our study assumes that it is fully rotationally induced and thus vanishes when the star's angular momentum goes to zero. The main advantage of the approach using the quadrupole parameter is that for NSs with a given EoS and mass, it does not depend on the NS rotational frequency, and, thus, it naturally scales $j$ and $q$. In terms of the spacetime geometry, its value determines whether the influence of the $t-\varphi$ component of the metric tensor and related frame-dragging effects dominate or are overtaken by the other components. Depending on the mass of the NS, the quadrupole parameter can range from $q/j^2~\gtrsim~1$ for a very massive (compact) NS up to $q/j^2~\sim~10$ for a low-mass NS \citep{1968ApJ...153..807H,2019ApJ...877...66U}. 

\begin{figure*}[t]
\centering
\includegraphics[width=1.
\linewidth]{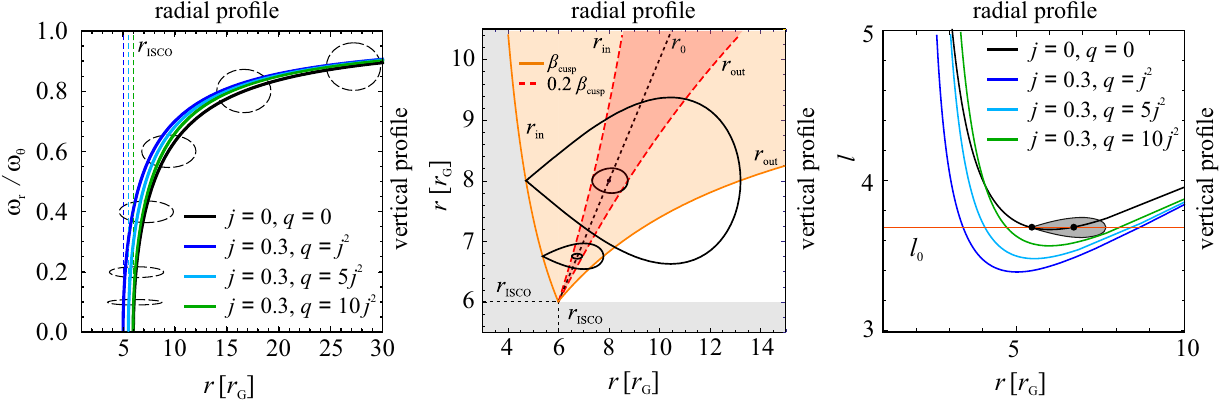}
\caption{{\emph{Left:}} Epicyclic frequencies ratio and shape of tori in the limit of $\beta\rightarrow 0$. The dashed-coloured vertical lines denote the locations of ISCO calculated for different spacetime parameters. The frequency-ratio values calculated for different spacetime parameters at given radii are indicated by the coloured curves and labelled by the standard (left and bottom) axes. Several additional dashed ellipses, drawn for a non-rotating star ($j=0$) and described by the upper and right axes, show the limiting cross-sections of slender tori located at a radius marked by the (common) lower axis.
{\emph{Middle:}} 
Radial extension and shape of a torus with its centre located at radius $r_0$  in the case of a non-rotating NS ($j=0$). The dependence of the inner ($r_{\mathrm{in}}$) and outer ($r_{\mathrm{out}}$) torus radius on $r_0$ is marked by orange ($\beta=\beta_{\mathrm{cusp}}$) and red ($\beta=0.2\,\beta_{\mathrm{cusp}}$) curves. Equipotential curves (black contours) are shown for two torus centre locations, $r_0=6.75\,\rg$ and $r_0=8\,\rg$. The torus radial profile coordinates are identical to those on the bottom axes.
{\emph{Right:}} Profiles of Keplerian specific angular momentum $l$ corresponding to the same spacetime parameters as in the left panel. The red horizontal line denotes a particular (chosen) value of $l_0$. The shape of the cusp torus corresponding to this value of $l_0$ and $j=0$ is shown. As in the middle panel, the torus radial profile coordinates are identical to those on the bottom axis. 
Shape of equipotentials for the same value of $l_0$ and same spacetime parameters are shown later, in Fig.~\ref{figure:eltoryL}.}
    \label{figure:torus_ratio}
\end{figure*}    

\section{Accretion tori around rotating neutron stars}
\label{Section:tori}

To calculate the accretion flow properties, we follow the Polish doughnut approach \citep[][]{Kozlowski-etal:1978:,1978A&A....63..221A,pac-abr:1982} and consider pressure-supported differentially rotating toroidal fluid flow embedded in the spacetime of a slowly rotating NS. 
The Polish doughnut model is frequently used to approximate a thick accretion disc around a compact object and agrees well with the results of numerical simulations. The equipotential surfaces given by the model mimic well a wide range of the properties of simulated accretion flows and other analytical models in both optically thin and thick regimes \citep{2009A&A...498..471Q,2017A&A...607A..68G,2019ApJ...884L..37L}.

\subsection{Polytropic tori}
\label{Section:poly}

To model the thermodynamic properties of the fluid, we adopt a polytropic approximation of a power-law relation between the gas pressure and the energy density. 
This approach gives excellent results in modelling compact stars' structure and provides a reasonable first probe into the complicated dynamics of accretion flows. Polytropic models were involved in original models of Polish doughnuts \citep{1978A&A....63..221A, Kozlowski-etal:1978:}, the first studies of oscillations and stability of geometrically thick accretion discs \citep{Papaloizou+Pringle1984, Blaes1985, 2006MNRAS.369.1235B}, calculations of the vertical structure of geometrically thin accretion discs \citep{2008bhad.book.....K, Kluzniak+Kita2000} and their trans-sonic nature \cite{Abramowicz+Zurek1981, Abramowicz+Kato1989}. Validity of the polytropic approximation in the case of cores of optically thin accretion discs was demonstrated by \citet{Ketsaris+Shakura1998}. A clear advantage of this approach is eliminating the energy equation from the discussion, which considerably simplifies the problem. A drawback is a significant limitation of the number of possible rotation laws that must obey the relativistic Von-Zeipel theorem \citep{Abramowicz1971}. 

Adopting this approach, we assume that both the spacetime and the flow share the same symmetries, i.e., they are axisymmetric and stationary. We neglect the self-gravity of the flow, leaving the central star as the only source of gravity.

The spacetime is described by the line element (\ref{eq:stel}). The four-velocity of the fluid then has only two non-zero components,
\begin{equation}
	u^\mu = A(1,0,0,\Omega),
\end{equation}
where $A=u^{t}$ is the redshift function and $\Omega=u^\varphi/u^t$ is the orbital velocity with respect to a distant observer. Functions $A$ and $\Omega$ are given by
\begin{align}
    A =& u^{t} = 
    \left(-g_{tt} - 2 \Omega g_{t\varphi} - \Omega^2 g_{\varphi\varphi}\right)^{-1/2}, 
    \\
	\Omega =& \dfrac{u^\varphi}{u^t} = 
	\frac{g^{t\varphi} - l g^{\varphi\varphi}}{g^{tt} - l g^{t\varphi}},
\end{align}
where $l$ denotes the specific angular momentum of the flow, $l=-u_\varphi/u_t$.

Perfect fluid of a rest-mass density $\rho$, pressure $p$ and total energy density $e$ is characterised by the stress-energy tensor
\begin{equation}
T^{\mu\nu} = \left(p+e\right) u^\mu u^\nu + p g^{\mu\nu}.
\end{equation}

For a polytropic fluid, we may write: 
\begin{align}
    \label{pol}
	p &= K \rho^{1+1/n}, 
	\\
	e &= np + \rho.
\end{align}
where $K$ and $n$ denote the polytropic constant and the polytropic index, respectively.

\subsection{Distribution of specific angular momentum and other quantities}
\label{ssec:Tori2}

The structure of the flow follows from the relativistic conservation laws,
\begin{equation}
\nabla_\mu T^\mu_{\; \ \; \nu} = 0.
\end{equation}
In the next, we assume the specific angular momentum to be constant inside the torus, $l=l_0$, where the subscript 0 refers to the torus centre ($r_0$). The Euler equation can be then simply written as  ~\citep{1978A&A....63..221A,2006CQGra..23.1689A} 
\begin{equation}
    \label{eul}
	\nabla_\mu \left(\ln \en\right) = - \frac{\nabla_\mu p}{p + e}, 
\end{equation}
with  $\en$  being the specific mechanical energy 
\begin{equation}
	\en = - u_t = \left( - g^{tt} + 2 l g^{t\varphi} - l^2 g^{\varphi\varphi} \right)^{-1/2}.
\end{equation}
It is related to the effective potential $\pot$ of a test particle with specific angular momentum $l$ by $\pot=\ln\en$ \citep{2005Ap&SS.300..127A}. In the case of a polytropic fluid, Eq.~(\ref{eul}) can be integrated to obtain an algebraic relation (the Bernoulli equation),  
\begin{equation}
    \label{ber}
    H \en = \mathrm{const}.,
\end{equation}
where $H = (p+e)/\rho$ denotes the specific enthalpy in the form introduced by \cite{2016MNRAS.461.1356F} and \cite{2017bhns.work...47H}.

Equation (\ref{ber}) determines the structure and shape of the torus with the centre located at $r_0$, where the orbital velocity of the flow is Keplerian,
\begin{equation}
\Omega\, |_{r = r_{0}} \equiv\Omega_{0}=\Omega_{\mathrm{K}}\,.
\end{equation}
The formulae for the pressure and density distributions can be written, with respect to their values at $r_0$, as 
\begin{align}
	p &= p_0 \left [ f \left( r,\theta \right ) \right ]^{n+1},  \\ 
\label{equation:rho}	
 \rho &= \rho _0 \left [ f \left( r,\theta \right ) \right ]^n,
\end{align}
where
\begin{equation}
    f\left( r, \theta \right) = 1 - \frac{1}{n\sound^2}  \left( 1 -   \frac{\en_0}{\en} \right)
    \label{eq:LE1}
\end{equation}
is the Lane-Emden function, and $\sound$ is the sound speed at the torus centre given by ~\citep{2006CQGra..23.1689A,2016MNRAS.461.1356F}
\begin{equation} 
 	\sound^2 =  \left(\frac{\partial p}{\partial e}\right)_{0}  = \frac{(n+1) p_0}{n \left[\rho_0 + \left(n+1\right) p_0 \right]}.
\end{equation}
The surface of the torus is given by the condition $p=0$ and thus $f(r, \theta ) =0$. The torus centre corresponds to $f=1$. Equation (\ref{eq:LE1}) implies that the surfaces of constant $f$ coincide with those of constant $\en$ and are, therefore, entirely given by the spacetime geometry and the value of the specific angular momentum. The particular value of the central sound speed $\sound$ affects only 
the relation $f(\en)$. It is useful to introduce the $\beta$~parameter determining the torus thickness, which is connected to $\sound$ in the following manner \citep{2006CQGra..23.1689A,2006MNRAS.369.1235B}:
\begin{align}
	\beta^2 =& \frac{2n\sound^2}{r_0^2 \Omega_0^2 A_0^2}.
 \label{equation:beta}
\end{align}

\begin{figure*}[t]
    \centering
    \includegraphics[width=1.\linewidth]{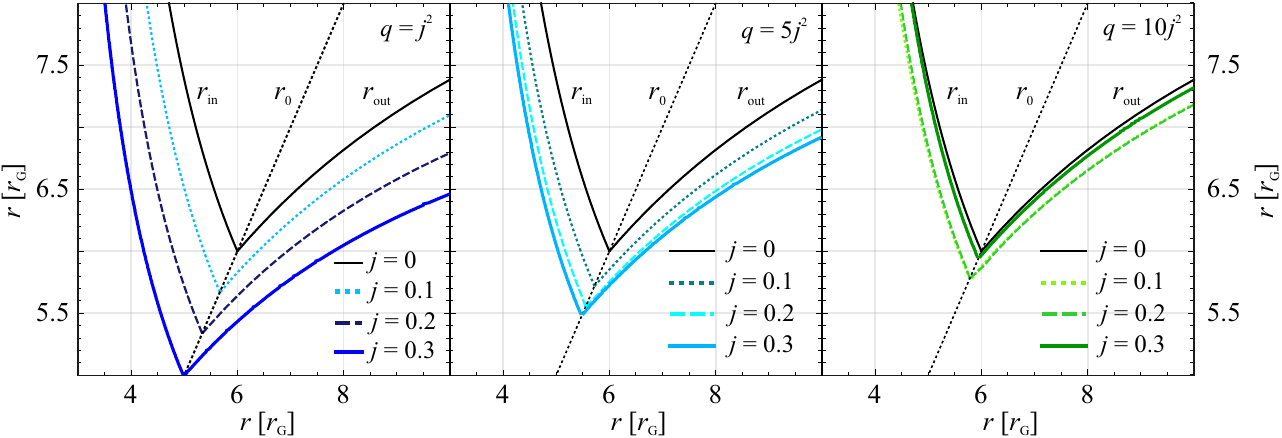}
    \caption{The location of the cusp ($r_\mathrm{in}$) and the outer radius ($r_\mathrm{out}$) of the cusp tori for different values of $r_0$ and spacetime parameters, increasing in $q/j^2$ from left to right.  The curves denoting inner and outer radii merge in the limit of the slender cusp torus, i.e., when $r_0=r_\mathrm{ISCO}$. The behaviour for the non-rotating case corresponds to the situation depicted along with several related details in the middle panel of Fig.~\ref{figure:torus_ratio}. 
    }
    
    \label{figure:jq}
\end{figure*}

\noindent By introducing the $\beta$~parameter, we eliminate the need to select specific values of $\sound$ and $n$ to further investigate tori and their shapes.

\subsection{Shape of tori}

The equipotential surfaces can be obtained using the quadratic Lane-Emden function as
\begin{equation}
    f^{(0)} = 1 - \bar{\omega}_r^2\bar{x}^2 - \bar{\omega}_\theta^2\bar{y}^2 = \mathrm{const.},
    \label{f0}
\end{equation}
where
\begin{align}
	\bar{x} =& \frac{\sqrt{ g_{rr,0} }}{\beta} \left( \frac{r - r_0}{r_0} \right)
 \\ \mathrm{and}\nonumber\\ 
	\bar{y} =& \frac{\sqrt{g_{\theta\theta,0}}}{\beta} \left( \frac{\frac{\pi}{2} - \theta}{r_0} \right)
\end{align}
are coordinates scaled by the radial end vertical thickness of the torus ($\bar{x}$ and $\bar{y}$ remains of the order of unity inside the torus even when $\beta\rightarrow 0$), and $\bar{\omega}_r$ and $\bar{\omega}_\theta$, given by 
\begin{equation}
    \bar{\omega}_X = \frac{\omega_{X0}}{\Omega_0},
    \quad
    \omega_{X0}^2 = \left(\frac{1}{A^2 g_{XX}}\frac{\partial^2 \ln\en}{\partial X^2}\right)_0
    \quad
    \left(X = r,\theta\right),
    \label{eq:epicyclic-frequencies}
\end{equation}
are normalised radial and vertical epicyclic frequencies defined at the centre of the torus \citep{2005Ap&SS.300..127A}, which are discussed in detail in our Paper II.

Filling up the largest closed equipotential surface results in the largest conceivable torus. Depending on the position of the torus centre $r_0$, one of the two cases may occur. First, for tori close enough to the central object, a critical value of $\beta$ parameter exists, $\beta_\mathrm{cusp}$, giving the largest equipotential implying a self-crossing surface with a cusp located in the equatorial plane between the innermost stable circular orbit (ISCO) and the innermost bound orbit (IBO). The cusp's location corresponds to the local maximum of the effective potential in the radial direction. Consequently, even a slight perturbation of such a torus with a cusp may lead to accretion that is only driven by pressure and inertial forces \citep{Kozlowski-etal:1978:}.

Conversely, for tori centred further away, the largest equipotential separates the closed ones from those open to infinity. The inner edge of such tori is located at the marginally bound orbit. However, before it may accrete onto the central compact object, the matter at the inner edge must first lose some angular momentum (likely through viscous processes). 

In the opposite situation, when $\beta\rightarrow 0$ (slender-torus limit), the torus' radial and vertical extensions become insignificant compared to its circumferential radius. Its poloidal cross-section then has an elliptical shape with radial and vertical semi-axes in the ratio inverse to the ratio of the vertical and radial epicyclic frequency  \citep[][]{2006CQGra..23.1689A, 2006MNRAS.369.1235B}. As $r_0$ approaches the ISCO, the ratio of epicyclic frequencies vanishes with vanishing radial epicyclic frequency. The detailed dependence of this ratio on the spacetime parameters across their astrophysically relevant range is illustrated in the left panel of Fig.~\ref{figure:torus_ratio}. It should be noted that since the influence of the quadrupole moment increases with increasing angular momentum, but at the same time is opposite, the behaviour corresponding to the moderate angular momentum along with the highest quadrupole moment considered ($j=0.3$, $q=10j^2$) is quite similar to the non-rotating case.

\subsection{Explicit formulae in Hartle-Thorne spacetime}

The above general formulae describing tori are given in a self-consistent form, allowing for a straightforward description of tori. The implied explicit formulae describing the full solution for the accretion tori in the Hartle-Thorne spacetime are rather long, and it would be challenging to fit them within this paper. Instead, we provide their complete version in a Wolfram Mathematica notebook\footnote{\url{https://github.com/Astrocomp/HTtori}}. This notebook contains functions for calculating important quantities of the metric and the shape of the equipotential surfaces for given parameters. The notebook includes a commentary to make it easier to read and use.

\section{{Solutions for astrophysically relevant spacetime parameters}}
\label{section:astro}

We illustrate the radial extension of tori around a non-rotating NS along with several equipotentials determining the shape of possible tori in the middle panel of Fig.~\ref{figure:torus_ratio}. In the case of the rotating NS, the spacetime parameters will affect the shape of the solution, especially as it depends on the Keplerian angular momentum profile  (as shown in the right panel of Fig.~\ref{figure:torus_ratio}.

\subsection{Impact of spacetime parameters on the shape of tori}

In the right panel of Fig.~\ref{figure:torus_ratio}, we demonstrate the behaviour of the Keplerian angular momentum for chosen specific spacetime parameters and compare it with the constant angular moment of the fluid (which is constant across the torus), and indicate the corresponding radial extension of the torus. In Fig.~\ref{figure:jq}, we show how the radial extension of the tori depends on the location of the torus centre for fixed values of $j$ and $q/j^2$. We can see that the influence of NS rotation is greater for small values of $q/j^2$ (left panel) and that, similarly to the case of the geodesic epicyclic frequency ratio, the behaviour corresponding to the moderate angular momentum along with the highest quadrupole moment considered, $j=0.3$ and $q=10j^2$, is rather similar to the non-rotating case (right panel).

\begin{figure*}[t]
    \centering
    \includegraphics[width=1.\linewidth]{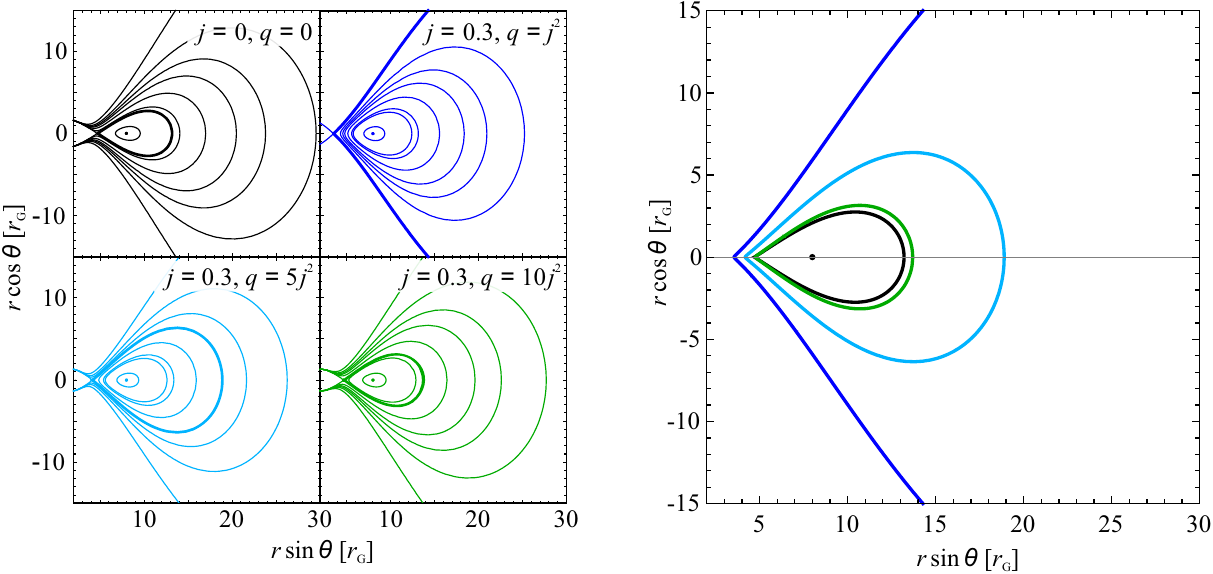}
    \caption{Shape of tori calculated for a fixed torus centre, $r_0$.  
    \emph{Left}: Shape of tori calculated for the set of spacetime parameters taken from Fig.~\ref{figure:torus_ratio} and the fixed value of $r_0=8r_{\mathrm{G}}$  (relevant values of $l_0$ are determined by spacetime parameters). The profiles are calculated for various increasing values of the thickness parameter $\beta$. These values are chosen to smoothly cover the range from $\beta=0$ corresponding to the case of a slender torus (central dot), through values corresponding to closed equipotential structures - tori, including the limiting case of cusp equipotential (cusp torus), to the values corresponding to open equipotentials. {\emph{Right}}: The cusp tori configurations from the left panel are compared. The black dot denotes the centre for all tori. 
    }
    
    \label{figure:eltoryR}
\end{figure*}

\begin{figure*}[t]
    \centering
    \includegraphics[width=1.\linewidth]{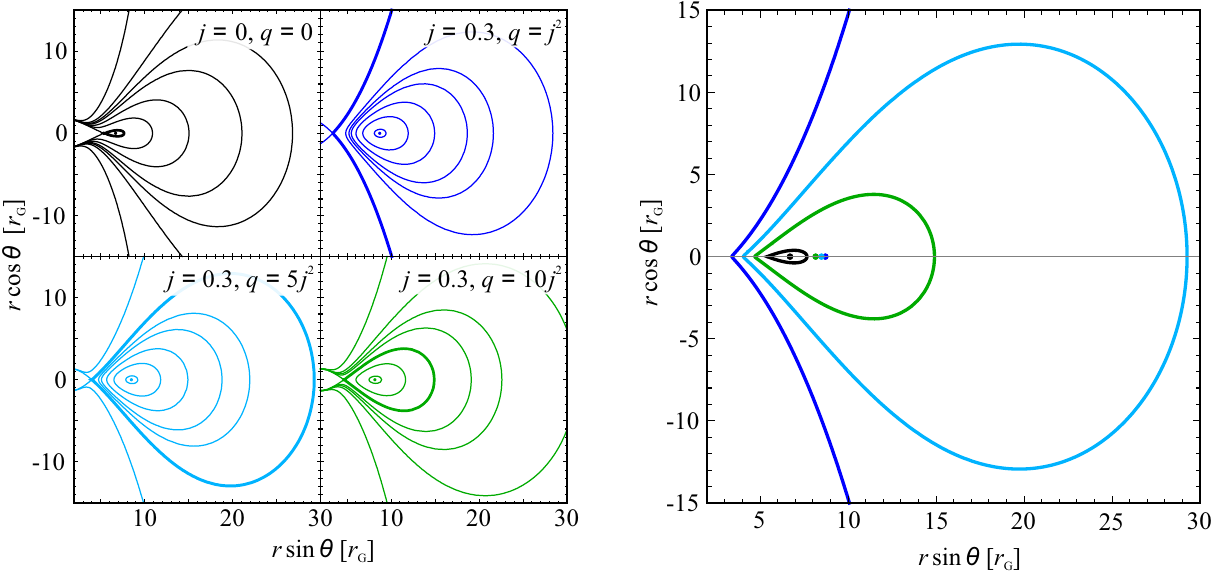}
    \caption{{The shape of tori calculated for a fixed value of the specific angular momentum $l_0$. \emph{Left}}: The case of the same set of spacetime parameters and a fixed value of $l_0$ as in Fig.~\ref{figure:torus_ratio}. The profiles are again determined for several values of the thickness parameter $\beta$. {\emph{Right}}: The cusp tori configurations from the left panel are compared. The dots denote the centres of each corresponding torus. 
    }
    \label{figure:eltoryL}
\end{figure*}

In Fig.~\ref{figure:eltoryR}, we present examples of the meridional cross-sections of the tori for several values of $j$ and $q/j^2$ in the case of fixed torus centre position, $r_0$. While the left panel shows several examples of various cross-sections, special attention is given to the cusp tori cases within the right panel.

We can see that, similar to the situation of slender tori determined by the ratio of the geodesic epicyclic frequencies depicted in the left panel of Fig.~\ref{figure:torus_ratio}, the combined effect of the NS's angular momentum and quadrupole moment on the non-slender torus shape does not imply any qualitative differences. Furthermore, the cusp tori corresponding to a moderate angular momentum ($j=0.3$) and a high quadrupole moment($q=10j^2$) are rather similar to the case of a non-rotating star. Indeed, the differences between the size of the cusp torus implied for $j=0$ (black curve) and that for $j=0.3,$ and $q=10j^2$ (green curve) are minimal.

\subsection{Coordinate-independent comparison}

In Fig.~\ref{figure:eltoryL}, we present examples of the meridional cross-sections in the case of a fixed value of $l_0$, which corresponds to the value marked in the right panel of Fig.~\ref{figure:torus_ratio}. Even in this case, the same effect of compensation between the impacts of the angular momentum and quadrupole moment is clearly present.  However, the differences between the size of the cusp torus implied for $j=0$ (black curve) and that for $j=0.3,$ and $q=10j^2$ (green curve) are much more apparent.

By integrating Eq.~(\ref{equation:rho}), we can compare the sizes of the two cusp tori calculated for the same value of $l_0$, denoted by the black and green contours in Fig.~\ref{figure:eltoryL}. We find that the torus calculated for the rotating oblate NS has a central density and mass of several orders of magnitude higher than that calculated for a non-rotating NS. 

\section{Discussion and conclusions}
\label{Section:conclusions}

We studied tori with thickness determined by the $\beta$ parameter and, utilising the quadrupole parameter, found that the influences of the NS's angular momentum and quadrupole moment were rather counterbalanced, i.e. that for the fixed radius of the centre, the solution for moderate angular momentum and relatively large but still plausible quadrupole parameter was similar to the non-rotating case. This finding is in accordance with the results of several recent studies \citep{kluzniakrosinska2013,kluzniakrosinska2014,rosinskaetal2014} which explored the interplay between the relativistic frame dragging effects given by the impact of angular momentum and the effects associated with the quadrupole moment that would also arise in Newtonian physics. Nevertheless, as we show in comparison between the non-rotating and moderately rotating cases, cusp tori with the same specific angular momentum have very different sizes for different choices of spacetime parameters.

\subsection{Influence of the NS EoS}
\label{Section:EOS}

So far, we have focused on examining the effects of spacetime parameters on accretion tori while considering a general range of these parameters associated with up-to-date models of NSs. However, a more thorough investigation of the implications given by the NS EoS is needed. In particular, the impact of the NS's radius represents a significant challenge.

The inner edge of a torus with a given central radius and thickness parameter must be above the NS surface. Thus, the existence of a torus depends on the NS EoS since it determines the NS radius assigned to a given mass and the rotation frequency. In this context, a necessary condition for the existence of cusp tori is that the NS surface is located below the ISCO, while the sufficient condition is that it is located below IBO \citep[e.g.,][]{Kozlowski-etal:1978:,1978A&A....63..221A,pac-abr:1982}.

\begin{figure}[t]
	\begin{center}
		\includegraphics[width=1.\linewidth]{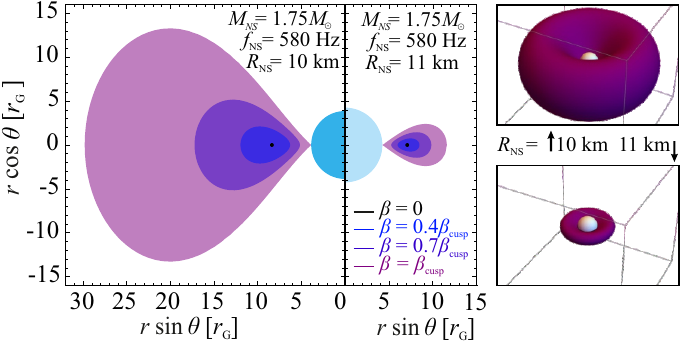}
		\caption{Sizes of two NSs of the same mass and rotational frequency but two different radii, calculated according to universal relations. The set of coloured tori illustrates the corresponding two different limits on the maximal size of the cusp tori that do not yet overlap with the star. The two NSs, along with their maximal cusp tori, are also shown in a 3D projection. \label{figure:EoSLT}}
  
	\end{center}
\end{figure}

The radii of NSs given by various EoS are well determined by recently identified universal relations \citep[][]{2013PhRvD..88b3007M,2013MNRAS.433.1903U,2013PhRvD..88b3009Y,2013Sci...341..365Y,2015MNRAS.454.4066P,2017MNRAS.470L..54R}. In Fig.~\ref{figure:EoSLT}, assuming these universal relations, we illustrate the size of two possible stars with the same mass and rotational frequency but different  NS angular momentum and quadrupole moment (namely, $ j = 0.205 $, $ q = 2.28 j^2$ and $ j = 0.236 $, $ q = 2.84 j^2$)\footnote{STL files with the 3D models of these two solutions can be downloaded from \url{https://www.printables.com/model/794643-tori-around-neutron-stars}.}. For each of these two stars, we find and label the surface of the largest possible cusp torus that does not overlap with the NS surface and the sizes of a torus with various thicknesses $\beta$. Clearly, while there is only a small difference in the NS size, the difference in the size of the tori is significant. Therefore, it is expected that the astrophysical applications of our results will be very sensitive to the NS EoS.

\subsection{Summary and overall implications}

In this work, we presented a solution of an equilibrium fluid tori in the Hartle-Thorne geometry describing the spacetime of a rotating NS. The solution includes the combined influence of the NS's angular momentum and quadrupole moment on the underlying properties of tori, such as the tori centre's position, its radial and vertical extent and its overall size. We have paid special attention to the critically thick configurations -- the cusp tori. 

Given the above discussion, we can conclude that the quadrupole moment induced by the NS rotation should influence the observable spectral and temporal behaviour of tori. The underlying astrophysical applications, including the exploration of observational signatures of the NS EoS, need to focus on specific physical phenomena (such as the QPOs discussed in Paper II) for which we have provided the publicly available tools to describe the tori. 

\begin{acknowledgements}
     We thank Marek Abramowicz and W{\l}odek Klu{\'z}niak for the valuable discussions. We also thank the reviewer for valuable comments and suggestions, which significantly helped to improve the paper. We acknowledge the Czech Science Foundation (GA\v{C}R) grant No.~21-06825X. The INTER-EXCELLENCE project No. LTT17003 is acknowledged by KK, and the INTER-EXCELLENCE project No. LTC18058 is acknowledged by MM and MU. We also thank the EU OPSRE project No. $\mathrm{CZ.02.2.69/0.0/0.0/18\_054/0014696}$ titled ``Development of R\&D capacities of the Silesian University in Opava'', and the PRODEX program of the European Space Agency (ref.\ 4000132152). VK acknowledges the Research Infrastructure LM2023047 of the Czech Ministry of Education, Youth and Sports. OS acknowledges funding by the Deutsche Forschungsgemeinschaft (DFG, German Research Foundation) under Germany's Excellence Strategy – EXC 2094 – 390783311. We furthermore acknowledge the support provided by the internal grants of Silesian University, $\mathrm{SGS/12,13/2019}$, $\mathrm{SGS/31/2023}$, $\mathrm{SGS/25/2024}$, and $\mathrm{SGF/1/2020}$, the latter of which has been carried out within the EU OPSRE project titled ``Improving the quality of the internal grant scheme of the Silesian University in Opava'', reg. number: $\mathrm{CZ.02.2.69/0.0/0.0/19\_073/0016951}$. 
\end{acknowledgements}
 
\bibliographystyle{aa}
\bibliography{references}

\newcommand{\noopsort}[1]{}
\begin{thebibliography}{63}
\expandafter\ifx\csname natexlab\endcsname\relax\def\natexlab#1{#1}\fi

\bibitem[{{Abramowicz} {et~al.}(1978){Abramowicz}, {Jaroszynski}, \& {Sikora}}]{1978A&A....63..221A}
{Abramowicz}, M., {Jaroszynski}, M., \& {Sikora}, M. 1978, \aap, 63, 221

\bibitem[{{Abramowicz}(1971)}]{Abramowicz1971}
{Abramowicz}, M.~A. 1971, \actaa, 21, 81

\bibitem[{{Abramowicz} {et~al.}(2003){Abramowicz}, {Almergren}, {Klu\'zniak}, \& {Thampan}}]{2003gr.qc....12070A}
{Abramowicz}, M.~A., {Almergren}, G.~J.~E., {Klu\'zniak}, W., \& {Thampan}, A.~V. 2003, arXiv e-prints, gr-qc/0312070

\bibitem[{{Abramowicz} {et~al.}(2006){Abramowicz}, {Blaes}, {Hor{\'a}k}, {Klu\'zniak}, \& {Rebusco}}]{2006CQGra..23.1689A}
{Abramowicz}, M.~A., {Blaes}, O.~M., {Hor{\'a}k}, J., {Klu\'zniak}, W., \& {Rebusco}, P. 2006, Classical and Quantum Gravity, 23, 1689

\bibitem[{{Abramowicz} \& {Kato}(1989)}]{Abramowicz+Kato1989}
{Abramowicz}, M.~A. \& {Kato}, S. 1989, \apj, 336, 304

\bibitem[{{Abramowicz} \& {Klu{\'z}niak}(2001)}]{2001A&A...374L..19A}
{Abramowicz}, M.~A. \& {Klu{\'z}niak}, W. 2001, \aap, 374, L19

\bibitem[{{Abramowicz} \& {Klu{\'z}niak}(2005)}]{2005Ap&SS.300..127A}
{Abramowicz}, M.~A. \& {Klu{\'z}niak}, W. 2005, \apss, 300, 127

\bibitem[{{Abramowicz} \& {Zurek}(1981)}]{Abramowicz+Zurek1981}
{Abramowicz}, M.~A. \& {Zurek}, W.~H. 1981, \apj, 246, 314

\bibitem[{{Blaes}(1985)}]{Blaes1985}
{Blaes}, O.~M. 1985, \mnras, 216, 553

\bibitem[{{Blaes} {et~al.}(2006){Blaes}, {Arras}, \& {Fragile}}]{2006MNRAS.369.1235B}
{Blaes}, O.~M., {Arras}, P., \& {Fragile}, P.~C. 2006, \mnras, 369, 1235

\bibitem[{{Bursa} {et~al.}(2004){Bursa}, {Abramowicz}, {Karas}, \& {Klu{\'z}niak}}]{bur-etal:2004:APJ}
{Bursa}, M., {Abramowicz}, M.~A., {Karas}, V., \& {Klu{\'z}niak}, W. 2004, APJL, 617, L45

\bibitem[{{Cackett} {et~al.}(2008){Cackett}, {Miller}, {Bhattacharyya}, {Grindlay}, {Homan}, {van der Klis}, {Miller}, {Strohmayer}, \& {Wijnands}}]{cac-etal:2008}
{Cackett}, E.~M., {Miller}, J.~M., {Bhattacharyya}, S., {et~al.} 2008, \apj, 674, 415

\bibitem[{{\noopsort{Cz}}{Č}ade\v{z} {et~al.}(2008){\noopsort{Cz}}{Č}ade\v{z}, {Calvani}, \& {Kosti{\'c}}}]{cad-etal:2008}
{\noopsort{Cz}}{Č}ade\v{z}, A., {Calvani}, M., \& {Kosti{\'c}}, U. 2008, A\&A, 487, 527

\bibitem[{{de Avellar} {et~al.}(2018){de Avellar}, {Porth}, {Younsi}, \& {Rezzolla}}]{2018MNRAS.474.3967D}
{de Avellar}, M. G.~B., {Porth}, O., {Younsi}, Z., \& {Rezzolla}, L. 2018, \mnras, 474, 3967

\bibitem[{{Fabian} {et~al.}(2000){Fabian}, {Iwasawa}, {Reynolds}, \& {Young}}]{rey-fab:2000}
{Fabian}, A.~C., {Iwasawa}, K., {Reynolds}, C.~S., \& {Young}, A.~J. 2000, \pasp, 112, 1145

\bibitem[{{Fabian} \& {Vaughan}(2003)}]{2003MNRAS.340L..28F}
{Fabian}, A.~C. \& {Vaughan}, S. 2003, \mnras, 340, L28

\bibitem[{{Fragile} {et~al.}(2016){Fragile}, {Straub}, \& {Blaes}}]{2016MNRAS.461.1356F}
{Fragile}, P.~C., {Straub}, O., \& {Blaes}, O. 2016, \mnras, 461, 1356

\bibitem[{{Frank} {et~al.}(2002){Frank}, {King}, \& {Raine}}]{2002apa..book.....F}
{Frank}, J., {King}, A., \& {Raine}, D.~J. 2002, {Accretion Power in Astrophysics: Third Edition} (Cambridge University Press)

\bibitem[{{Gimeno-Soler} \& {Font}(2017)}]{2017A&A...607A..68G}
{Gimeno-Soler}, S. \& {Font}, J.~A. 2017, \aap, 607, A68

\bibitem[{{Gondek-Rosi{\'n}ska} {et~al.}(2014){Gondek-Rosi{\'n}ska}, {Klu{\'z}niak}, {Stergioulas}, \& {Wi{\'s}niewicz}}]{rosinskaetal2014}
{Gondek-Rosi{\'n}ska}, D., {Klu{\'z}niak}, W., {Stergioulas}, N., \& {Wi{\'s}niewicz}, M. 2014, \prd, 89, 104001

\bibitem[{{Hartle} \& {Thorne}(1968)}]{1968ApJ...153..807H}
{Hartle}, J.~B. \& {Thorne}, K.~S. 1968, \apj, 153, 807

\bibitem[{{Hor{\'a}k} {et~al.}(2017){Hor{\'a}k}, {Straub}, {{\v{S}}r{\'a}mkov{\'a}}, {Goluchov{\'a}}, \& {T{\"o}r{\"o}k}}]{2017bhns.work...47H}
{Hor{\'a}k}, J., {Straub}, O., {{\v{S}}r{\'a}mkov{\'a}}, E., {Goluchov{\'a}}, K., \& {T{\"o}r{\"o}k}, G. 2017, in RAGtime 17-19: Workshops on Black Holes and Neutron Stars, 47--59

\bibitem[{{Ingram} \& {Done}(2010)}]{Ingram+Done:2010}
{Ingram}, A. \& {Done}, C. 2010, MNRAS, 405, 2447

\bibitem[{{Kato}(2001)}]{kat:2001}
{Kato}, S. 2001, PASJ, 53, 1

\bibitem[{{Kato} {et~al.}(2008){Kato}, {Fukue}, \& {Mineshige}}]{2008bhad.book.....K}
{Kato}, S., {Fukue}, J., \& {Mineshige}, S. 2008, {Black-Hole Accretion Disks --- Towards a New Paradigm ---} (Kyoto University Press)

\bibitem[{{Ketsaris} \& {Shakura}(1998)}]{Ketsaris+Shakura1998}
{Ketsaris}, N.~A. \& {Shakura}, N.~I. 1998, Astronomical and Astrophysical Transactions, 15, 193

\bibitem[{{Kluzniak} \& {Kita}(2000)}]{Kluzniak+Kita2000}
{Kluzniak}, W. \& {Kita}, D. 2000, arXiv e-prints, astro-ph/0006266

\bibitem[{{Klu{\'z}niak} \& {Rosi{\'n}ska}(2013)}]{kluzniakrosinska2013}
{Klu{\'z}niak}, W. \& {Rosi{\'n}ska}, D. 2013, \mnras, 434, 2825

\bibitem[{{Klu{\'z}niak} \& {Rosi{\'n}ska}(2014)}]{kluzniakrosinska2014}
{Klu{\'z}niak}, W. \& {Rosi{\'n}ska}, D. 2014, in Journal of Physics Conference Series, Vol. 496, Journal of Physics Conference Series, 012016

\bibitem[{{Kozlowski} {et~al.}(1978){Kozlowski}, {Jaroszynski}, \& {Abramowicz}}]{Kozlowski-etal:1978:}
{Kozlowski}, M., {Jaroszynski}, M., \& {Abramowicz}, M.~A. 1978, \aap, 63, 209

\bibitem[{{Lan{\v{c}}ov{\'a}} {et~al.}(2019){Lan{\v{c}}ov{\'a}}, {Abarca}, {Klu{\'z}niak}, {Wielgus}, {S{\k{a}}dowski}, {Narayan}, {Schee}, {T{\"o}r{\"o}k}, \& {Abramowicz}}]{2019ApJ...884L..37L}
{Lan{\v{c}}ov{\'a}}, D., {Abarca}, D., {Klu{\'z}niak}, W., {et~al.} 2019, \apjl, 884, L37

\bibitem[{{Lattimer} \& {Prakash}(2001)}]{Lattimer+Prakash:2001}
{Lattimer}, J.~M. \& {Prakash}, M. 2001, \apj, 550, 426

\bibitem[{{Lattimer} \& {Prakash}(2007)}]{2007PhR...442..109L}
{Lattimer}, J.~M. \& {Prakash}, M. 2007, \physrep, 442, 109

\bibitem[{{Lewin} \& {van der Klis}(2010)}]{lewin-etal:2010:book}
{Lewin}, W. \& {van der Klis}, M. 2010, {Compact Stellar X-ray Sources} (Cambridge University Press)

\bibitem[{{Lo} \& {Lin}(2011)}]{2011ApJ...728...12L}
{Lo}, K.-W. \& {Lin}, L.-M. 2011, \apj, 728, 12

\bibitem[{{Maselli} {et~al.}(2013){Maselli}, {Cardoso}, {Ferrari}, {Gualtieri}, \& {Pani}}]{2013PhRvD..88b3007M}
{Maselli}, A., {Cardoso}, V., {Ferrari}, V., {Gualtieri}, L., \& {Pani}, P. 2013, \prd, 88, 023007

\bibitem[{{Matuszkov{\'a}} {et~al.}(2024){Matuszkov{\'a}}, {T{\"o}r{\"o}k}, {Klimovi{\v{c}}ov{\'a}}, {Hor{\'a}k}, {Straub}, {{\v{S}}r{\'a}mkov{\'a}}, {Lan{\v{c}}ov{\'a}}, {Urbanec}, {Urbancov{\'a}}, \& {Karas}}]{paperII-key}
{Matuszkov{\'a}}, M., {T{\"o}r{\"o}k}, G., {Klimovi{\v{c}}ov{\'a}}, K., {et~al.} 2024, arXiv e-prints, arXiv:2403.16231, accepted for publication in A\&A

\bibitem[{{McClintock} {et~al.}(2011){McClintock}, {Narayan}, \& {Shafee}}]{2011blho.book..252M}
{McClintock}, J.~E., {Narayan}, R., \& {Shafee}, R. 2011, in Black Holes, ed. M.~{Livio} \& A.~M. {Koekemoer} (Cambridge University Press), 252--260

\bibitem[{{McClintock} \& {Remillard}(2006)}]{2006csxs.book..157M}
{McClintock}, J.~E. \& {Remillard}, R.~A. 2006, in Compact stellar X-ray sources, Vol.~39 (Cambridge University Press), 157--213

\bibitem[{{Miller} {et~al.}(2020){Miller}, {Zoghbi}, {Raymond}, {Balakrishnan}, {Brenneman}, {Cackett}, {Draghis}, {Fabian}, {Gallo}, {Kaastra}, {Kallman}, {Kammoun}, {Motta}, {Proga}, {Reynolds}, \& {Trueba}}]{mil-etal:2020}
{Miller}, J.~M., {Zoghbi}, A., {Raymond}, J., {et~al.} 2020, \apj, 904, 30

\bibitem[{{Mishra} {et~al.}(2017){Mishra}, {Vincent}, {Manousakis}, {Fragile}, {Paumard}, \& {Klu{\'z}niak}}]{mis-etal:2017:}
{Mishra}, B., {Vincent}, F.~H., {Manousakis}, A., {et~al.} 2017, MNRAS, 467, 4036

\bibitem[{{Paczy\'nski} \& {Abramowicz}(1982)}]{pac-abr:1982}
{Paczy\'nski}, B. \& {Abramowicz}, M.~A. 1982, APJ, 253, 897

\bibitem[{{Papaloizou} \& {Pringle}(1984)}]{Papaloizou+Pringle1984}
{Papaloizou}, J.~C.~B. \& {Pringle}, J.~E. 1984, \mnras, 208, 721

\bibitem[{{Pappas}(2015)}]{2015MNRAS.454.4066P}
{Pappas}, G. 2015, \mnras, 454, 4066

\bibitem[{{Piraino} {et~al.}(2012){Piraino}, {Santangelo}, {Kaaret}, {M{\"u}ck}, {D'A{\`\i}}, {Di Salvo}, {Iaria}, {Robba}, {Burderi}, \& {Egron}}]{pir-etal:2012}
{Piraino}, S., {Santangelo}, A., {Kaaret}, P., {et~al.} 2012, \aap, 542, L27

\bibitem[{{Qian} {et~al.}(2009){Qian}, {Abramowicz}, {Fragile}, {Hor{\'a}k}, {Machida}, \& {Straub}}]{2009A&A...498..471Q}
{Qian}, L., {Abramowicz}, M.~A., {Fragile}, P.~C., {et~al.} 2009, \aap, 498, 471

\bibitem[{{Reina} {et~al.}(2017){Reina}, {Sanchis-Gual}, {Vera}, \& {Font}}]{2017MNRAS.470L..54R}
{Reina}, B., {Sanchis-Gual}, N., {Vera}, R., \& {Font}, J.~A. 2017, \mnras, 470, L54

\bibitem[{{Reynolds} \& {Fabian}(2008)}]{rey-fab:2008}
{Reynolds}, C.~S. \& {Fabian}, A.~C. 2008, \apj, 675, 1048

\bibitem[{{Rezzolla} {et~al.}(2003{\natexlab{a}}){Rezzolla}, {Yoshida}, {Maccarone}, \& {Zanotti}}]{2003MNRAS.344L..37R}
{Rezzolla}, L., {Yoshida}, S., {Maccarone}, T.~J., \& {Zanotti}, O. 2003{\natexlab{a}}, \mnras, 344, L37

\bibitem[{{Rezzolla} {et~al.}(2003{\natexlab{b}}){Rezzolla}, {Yoshida}, \& {Zanotti}}]{rez-etal:2003}
{Rezzolla}, L., {Yoshida}, S., \& {Zanotti}, O. 2003{\natexlab{b}}, MNRAS, 344, 978

\bibitem[{{Riaz} {et~al.}(2020{\natexlab{a}}){Riaz}, {Ayzenberg}, {Bambi}, \& {Nampalliwar}}]{2020ApJ...895...61R}
{Riaz}, S., {Ayzenberg}, D., {Bambi}, C., \& {Nampalliwar}, S. 2020{\natexlab{a}}, \apj, 895, 61

\bibitem[{{Riaz} {et~al.}(2020{\natexlab{b}}){Riaz}, {Ayzenberg}, {Bambi}, \& {Nampalliwar}}]{2020MNRAS.491..417R}
{Riaz}, S., {Ayzenberg}, D., {Bambi}, C., \& {Nampalliwar}, S. 2020{\natexlab{b}}, \mnras, 491, 417

\bibitem[{{Shafee} {et~al.}(2008){Shafee}, {Narayan}, \& {McClintock}}]{2008ApJ...676..549S}
{Shafee}, R., {Narayan}, R., \& {McClintock}, J.~E. 2008, \apj, 676, 549

\bibitem[{{Steiner} {et~al.}(2010){Steiner}, {McClintock}, {Remillard}, {Gou}, {Yamada}, \& {Narayan}}]{2010ApJ...718L.117S}
{Steiner}, J.~F., {McClintock}, J.~E., {Remillard}, R.~A., {et~al.} 2010, \apjl, 718, L117

\bibitem[{{Steiner} {et~al.}(2009){Steiner}, {McClintock}, {Remillard}, {Narayan}, \& {Gou}}]{2009ApJ...701L..83S}
{Steiner}, J.~F., {McClintock}, J.~E., {Remillard}, R.~A., {Narayan}, R., \& {Gou}, L. 2009, \apjl, 701, L83

\bibitem[{{Stella} \& {Vietri}(1998)}]{ste-vie:1998a}
{Stella}, L. \& {Vietri}, M. 1998, in 19th Texas Symposium on Relativistic Astrophysics and Cosmology, ed. J.~{Paul}, T.~{Montmerle}, \& E.~{Aubourg}

\bibitem[{{Strohmayer} {et~al.}(1996){Strohmayer}, {Zhang}, {Swank}, {Smale}, {Titarchuk}, {Day}, \& {Lee}}]{1996ApJ...469L...9S}
{Strohmayer}, T.~E., {Zhang}, W., {Swank}, J.~H., {et~al.} 1996, \apjl, 469, L9

\bibitem[{{T{\"o}r{\"o}k} {et~al.}(2005){T{\"o}r{\"o}k}, {Abramowicz}, {Klu{\'z}niak}, \& {Stuchl{\'\i}k}}]{2005A&A...436....1T}
{T{\"o}r{\"o}k}, G., {Abramowicz}, M.~A., {Klu{\'z}niak}, W., \& {Stuchl{\'\i}k}, Z. 2005, \aap, 436, 1

\bibitem[{{Urbancov{\'a}} {et~al.}(2019){Urbancov{\'a}}, {Urbanec}, {T{\"o}r{\"o}k}, {Stuchl{\'\i}k}, {Blaschke}, \& {Miller}}]{2019ApJ...877...66U}
{Urbancov{\'a}}, G., {Urbanec}, M., {T{\"o}r{\"o}k}, G., {et~al.} 2019, \apj, 877, 66

\bibitem[{{Urbanec} {et~al.}(2013){Urbanec}, {Miller}, \& {Stuchl{\'\i}k}}]{2013MNRAS.433.1903U}
{Urbanec}, M., {Miller}, J.~C., \& {Stuchl{\'\i}k}, Z. 2013, \mnras, 433, 1903

\bibitem[{{Wagoner}(1999)}]{wag:1999}
{Wagoner}, R.~V. 1999, Physics Reports, 311, 259

\bibitem[{{Yagi} \& {Yunes}(2013{\natexlab{a}})}]{2013PhRvD..88b3009Y}
{Yagi}, K. \& {Yunes}, N. 2013{\natexlab{a}}, \prd, 88, 023009

\bibitem[{{Yagi} \& {Yunes}(2013{\natexlab{b}})}]{2013Sci...341..365Y}
{Yagi}, K. \& {Yunes}, N. 2013{\natexlab{b}}, Science, 341, 365

\end{thebibliography}

\end{document}